\documentstyle[preprint,aps,eqsecnum]{revtex}

\newcommand{\lbar}{\overline}
\newcommand{\spa}{\vspace{.25cm}}
\newcommand{\beq}{\begin{equation}}
\newcommand{\eeq}{\end{equation}}
\newcommand{\beqs}{\begin{eqnarray}}
\newcommand{\eeqs}{\end{eqnarray}}
\newcommand{\bary}{\begin{array}}
\newcommand{\eary}{\end{array}}

\newcommand{\e}{e}

\newcommand{\eV}{\mbox{\,{\rm eV}}}
\newcommand{\BR}{\mbox{\,{\rm BR}}}
\newcommand{\SUL}{\stackrel{{}_{SU(2)_L}}{\Longleftrightarrow}}
\newcommand{\mmbar}{M \overline M}
\newcommand{\mue}{\mu \to 3\,e}
\newcommand{\taumue}{\tau \to \mu \, e \, e}
\newcommand{\taumuec}{\tau^- \to \mu^+ \, e^- \, e^-}

\newcommand{\btau}{\mbox{$\tau$}}
\newcommand{\bDelta}{\mbox{$\Delta$}}

\newcommand{\picwidth}{25}        
\newcommand{\picheightLRSM}{8.0}  
\newcommand{\picheightSUSY}{8.5}  
\newcommand{\piclength}{0.6cm}    
\newcommand{\picdist}{2.5cm}      

\def\lsim{\ \rlap{\raise 3pt \hbox{$<$}}{\lower 3pt \hbox{$\sim$}}\ }

\def\a{\alpha}
\def\b{\beta}

\def\O{{\cal O}}

\def\SM{Standard Model}
\def\NP{New Physics}
\def\lfv{lepton flavor violation}
\def\lfvg{lepton flavor violating}
\def\AN{atmospheric neutrino}
\def\SN{solar neutrino}
\def\amd{$\mu^+ \to e^+ \, \bar\nu_e \, \nu_\ell$}
\def\amdtau{$\mu^+ \to e^+ \, \bar\nu_\tau \, \nu_\ell$}
\def\amdlv{$\mu^+ \to e^+ \, \bar\nu_e \, \bar\nu_\ell$}

\def\npb#1{Nucl.\ Phys.\ {\bf B#1}}
\def\plb#1{Phys.\ Lett.\ {\bf B#1}}
\def\prd#1{Phys.\ Rev.\ {\bf D#1}}
\def\prl#1{Phys.\ Rev.\ Lett. {\bf#1}}

\def\zpc#1{Z.~Phys.\ {\bf C#1}}
\def\epjc#1{Eur.~Phys.~J.\ {\bf C#1}}

\def\ijmpa#1{Int.\ J.\ Mod.\ Phys.\ {\bf A#1}}

\def\mpla#1{Mod. Phys. Lett. A {\bf #1}}

\begin{document}
\draft
{\tighten
\preprint{\vbox{\hbox{WIS-98/24/Sep-DPP}
                \hbox{SLAC-PUB-7950}
                \hbox{hep-ph/9809524}}}

\title{~ \\ Can lepton flavor violating interactions \\ 
            explain the LSND results?}
\footnotetext{\scriptsize Research at SLAC is supported by the 
                          U.S. Department of Energy 
                          under contract DE-AC03-76SF00515}
\author{Sven Bergmann\,$^a$ and Yuval Grossman\,$^b$}
\address{ \vbox{\vskip 0.truecm}
  $^a$Department of Particle Physics \\
  Weizmann Institute of Science, Rehovot 76100, Israel \\
\vbox{\vskip 0.truecm}
  $^b$Stanford Linear Accelerator Center \\
  Stanford University, Stanford, CA 94309}

\maketitle

\begin{abstract}
If the atmospheric and the solar neutrino problem are both explained
by neutrino oscillations, and if there are only three light neutrinos,
then all mass-squared differences between the neutrinos are known. In
such a case, existing terrestrial neutrino oscillation experiments
cannot be significantly affected by neutrino oscillations, but, in
principle there could be an anomaly in the neutrino flux due to new
neutrino interactions. We discuss how a non-standard muon decay \amd\
would modify the neutrino production processes of these experiments.
Since $SU(2)_L$ violation is small for New Physics above the weak
scale one can use related flavor-violating charged lepton processes to
constrain these decays in a model independent way.  We show that the
upper bounds on $\mu \to 3e$, muonium-antimuonium conversion and
$\tau \to \mu \, e \, e$ rule out any observable effect for the
present experiments due to \amd\ for $\ell=e,\mu,\tau$,
respectively. Applying similar arguments to flavor-changing
semi-leptonic reactions we exclude the possibility that the
``oscillation signals'' observed at LSND are due to flavor-changing
interactions that conserve total lepton number.

\end{abstract} 
} 

\newpage


\section{Introduction}

There are strong experimental hints that suggest that the neutrino
sector is more complicated than it is in the \SM. In particular, the
\AN\ anomaly~\cite{AN} and the \SN\ problem~\cite{SN} can be explained
with {\it massive} neutrinos.

The \AN\ anomaly~\cite{AN} is the observation that the ratio of muon
neutrinos to electron neutrinos that are produced in the atmosphere is
about 0.6 of the theoretical expectation assuming \SM\ neutrinos.
Recently, the Super-Kamiokande Collaboration has
published~\cite{ANSuperK98} the analysis of their atmospheric neutrino
data from a 33.0 kiloton-year (535-day) exposure. The data exhibit a
zenith angle dependent deficit of muon neutrinos which cannot be
explained with the \SM\ massless neutrinos.  The estimated probability
that the observed $\mu/e$ ratio could be due to statistical
fluctuations is less than $10^{-5}$ (for the sub-GeV data), which is
widely considered as the first ``proof'' for massive neutrinos. The
data are consistent with~\cite{ANSuperK98}:
\beq \label{ANrange}
5 \times 10^{-4} \eV^2 < \Delta m^2 < 6 \times 10^{-3} \eV^2 \qquad
\sin^2 2\theta > 0.82 ~~~(90\%~\mbox{\rm C.L.}),
\eeq
where $\Delta m^2$ is the mass-squared difference and $\theta$ is
vacuum mixing angle for the favored $\nu_\mu \leftrightarrow \nu_\tau$
oscillations.  Note that $\nu_\mu \leftrightarrow \nu_e$ oscillations 
are disfavored by the observed zenith angle distribution and by the fact
that the up-to-down ratio for $\nu_\mu$-induced events departs much
more from unity than for the $\nu_e$-induced events.  Moreover the
CHOOZ experiment~\cite{CHOOZ} independently rules out $\bar \nu_e 
\leftrightarrow \bar \nu_\mu$ oscillation for mixing as 
large as in~(\ref{ANrange}) and $\Delta m^2 > 10^{-3} \eV^2$.

The long standing \SN\ puzzle~\cite{SN} is now confirmed by five
experiments using three different experimental techniques
and thus probing different neutrino energy ranges. All these
experiments observe a solar neutrino flux that is smaller than
expected. The most plausible solution is that the neutrinos are
massive and there is mixing in the lepton sector. Then neutrino
oscillations can explain the deficit of observed neutrinos with
respect to the Standard Solar Model. In the case of matter-enhanced
neutrino oscillations, the famous MSW effect provides an elegant
solution to the solar neutrino problem. The best fit is obtained for
the {\it small angle solution} which is given by \cite{SN}
\beq \label{SNrange} 
\Delta m^2= 5.4\times 10^{-6} \eV^2, \qquad 
\sin^2 2\theta = 6.0 \times 10^{-3}, 
\eeq 
where $\theta$ is vacuum mixing angle in a two {\it active} neutrino
framework involving the $\nu_e$ and either $\nu_\mu$ or $\nu_\tau$.
The {\it large angle solution} can also explain the data (with a worse
fit) with $\Delta m^2 = 1.8 \times 10^{-5} \eV^2$ and $\sin^2 2\theta
= 0.76 \,$.  Finally, {\it vacuum oscillations} provide an alternative
solution with the best-fit solution~\cite{SN} given by $\Delta m^2 =
8.0 \times 10^{-11} \eV^2$ and $\sin^2 2\theta = 0.75\,$.

It is well known that the \SM\ contains only three generations of
neutrinos and that SLD and LEP data exclude the existence of a fourth
light sequential neutrino \cite{PDG}. If, indeed, there are only three
light neutrinos, then an important consequence of the above solutions
(\ref{ANrange}) and (\ref{SNrange}) to the two different neutrino
anomalies is that all light neutrino mass-squared differences $\Delta
m^2_{ij} \equiv m_i^2-m_j^2$ are completely determined. The reason is
that with three generations, there are only two independent mass
differences since $\Delta m^2_{21} + \Delta m^2_{32} = \Delta
m^2_{31}$.  In particular, we learn that for any $i,j=1,2,3,$
\beq \label{terange} 
\Delta m^2_{ij} \lsim 10^{-2}~\eV^2.
\eeq 
This is below the sensitivity of all existing terrestrial experiments
(except the above mentioned CHOOZ experiment~\cite{CHOOZ}, which
provides an even stronger bound than (\ref{terange}) for $e-\mu$
oscillations and large mixing).  The conclusion is that if both
the atmospheric neutrino anomaly and solar neutrino problem are
explained by neutrino oscillations and there are only three light
neutrinos, then an extended three generation \SM, which allows for
small neutrino masses but leaves all interactions as they are in the
\SM, predicts that no anomaly should be observed in any terrestrial
neutrino experiment.
 
In contrast to this expectation, the LSND collaboration has reported a
positive signal in two different appearance channels. The first
analysis~\cite{LSND-DAR} uses $\bar \nu_\mu$'s from muon decay at rest
(DAR) and searches for $\bar \nu_e$'s via inverse beta decay. The
observed excess of $\bar \nu_e$ events corresponds to
an average transition probability of~\cite{LSND-DAR}
\beq \label{LSNDdata}
P(\bar \nu_\mu \to \bar \nu_e)=(3.1^{+1.1}_{-1.0} \pm 0.5) \times 10^{-3}.
\eeq
This result by itself could be explained by neutrino oscillations,
with $\Delta m^2$ and $\sin^2 2\theta$ in the range indicated in
Fig.~3 of Ref.~\cite{LSND-DAR}. Taking into account the restrictions
from the null results of other experiments, the preferred values of
the neutrino parameters are $\Delta m^2 \approx 2~\eV^2$ and $\sin^2
2\theta \approx 2 \times 10^{-3}$ and the {\it lower} limit on $\Delta
m^2$ for the neutrino oscillation solution is given by
\beq \label{LSNDrange} 
\Delta m^2 > 0.3~\eV^2.
\eeq 
The second analysis~\cite{LSND-DIF} uses $\nu_\mu$'s from pion decay
in flight (DIF) and searches for $\nu_e$'s via the $\nu_e \, C \to e^-
\, X$ inclusive reaction. Again, a positive signal has been reported,
which could be explained with neutrino oscillations that require
neutrino parameters similar to those of the DAR result. However,
the statistical significance of this result is much smaller than the
one of the DAR analysis.

Obviously, the lower bound (\ref{LSNDrange}) on $\Delta m^2$ is
incompatible with the neutrino oscillation solutions to the \AN\
anomaly (\ref{ANrange}) and the \SN\ problem (\ref{SNrange}) in a
three generation framework. One possibility is to postulate a light
``sterile neutrino''~\cite{ster,natsinglet}: a \SM\ singlet that mixes
with the active neutrinos. Then there would be four neutrino masses
which give three independent mass differences, as required to explain
the three mentioned results~\cite{Barger4nu}. Although adding ad hoc
this sterile neutrino would be phenomenologically satisfactory, it is
not well-motivated to have a light SM singlet.  (For attempts to 
naturally get a light sterile neutrino see e.g., \cite{natsinglet}.)

Due to the unappealing theoretical feature of a light sterile
neutrino, it is interesting to look for alternatives that could
explain the LSND anomaly with the known three light neutrinos only.
The authors of Ref.~\cite{Ful} have suggested that the \AN\ anomaly
and the LSND result are explained by the same mass-squared
difference. In Ref.~\cite{Pakvasa} a scenario where both the \SN\ and
the \AN\ anomalies are solved by the same $\Delta m^2$ has been
studied.  While these explanations were marginally consistent at the
time, they are excluded by the latest data. (There has been another
more recent attempt~\cite{Thun} to explain all experimental data
except the Homestake measurement with three active neutrinos
only. However their results have been criticized by the authors
of~\cite{FogliAN}.)

The aim of this work is to investigate another approach. We assume
that the three light neutrinos are not only massive but also interact
through \lfvg\ interactions, which are forbidden in the \SM. (We do not
consider here interaction that violate total lepton number, which will 
be studied separately~\cite{tlnv}.) This is an attractive possibility,
because various extensions of the \SM\ which predict neutrino masses
also give rise to such new interactions. These interactions can affect
the LSND production or detection processes or both.  We analyze the
consequences of small \lfvg\ interactions for short-baseline neutrino
oscillations experiments like LSND under the assumption that all
neutrino parameters are fixed to solve the \AN\ and the \SN\ anomaly.
We find that such a scenario, where new interactions explain the LSND
result(s), can be ruled out in a model independent way.

We note that the implications of exotic muon decays on the LSND
neutrino production have been studied by Herczeg~\cite{Herczeg}
showing within two explicit models, the left-right symmetric model
(LRSM) and SUSY without $R$-parity, that new interactions 
are too small to be relevant for LSND. More recently the
authors of Ref.~\cite{JM} have argued in favor of such a solution
(claiming that the DAR result could be explained within
LRSMs). However, they seem to have overlooked the strongest
experimental bound coming from muonium-antimuonium conversion.

Our paper is organized as follows.  In Section~\ref{lfvmd} we
introduce the formalism to describe the flavor violating interactions.
In Section~\ref{expbounds} we present the experimental bounds on
$SU(2)_L$ related \lfvg\ interactions containing only charged leptons.
In Section~\ref{specmodel} we show how these bounds can be used to
derive constraints on \amd\ within specific extensions of the \SM. We
generalize this idea in Section~\ref{modelind} and show in a model
independent way that the anomalous muon decay cannot have a detectable
effect in existing terrestrial neutrino oscillation experiments.  In
Section~\ref{semileptonic} we extend our analysis to semi-leptonic
reactions and argue that also in this case the bounds on $SU(2)_L$
related processes involving only charged fermions can be used to rule
out model independently the possibility that \lfvg\ interaction which
conserve total lepton number provide a valid explanation for the LSND
results.  We conclude in Section~\ref{conclusion}.


\section{Neutrino oscillations and New Interactions}
\label{lfvmd}

We start by reviewing the formalism of oscillation experiments in the
presence of non-standard neutrino interactions~\cite{yuval}.  To
illustrate this ``hybrid'' situation of having both non-trivial
neutrino properties and new interactions, we assume two neutrino
flavors, CP conservation, that the new interactions have the same Dirac
structure as the standard one and that the neutrinos are highly
relativistic. In general, in the presence of \NP, the neutrinos that
are produced and detected are not the weak eigenstates. Therefore, we
denote these neutrino states by the super-indices $p$ and $d$ which
stand for {\it production} and {\it detection}, respectively.
Consider the LSND setting: Anti-neutrinos are produced by $\mu^+\to
e^+\nu_e^p \bar\nu_\mu^p \,$, and detected by $\bar \nu_e^d + p \to
e^+ + n$.  We define the relevant mixing angles
\beq
\sin\theta_{pd} \equiv \langle \bar\nu_\mu^p | \bar\nu_e^d \rangle,\qquad
\sin\theta_{md} \equiv \langle \bar\nu_2 | \bar\nu_e^d \rangle, \qquad
\sin\theta_{mp} \equiv \langle \bar\nu_1 | \bar\nu_\mu^p \rangle,
\eeq
where $\nu_1$ and $\nu_2$ are mass eigenstates.  Then in the presence
of \lfvg\ interactions, the probability of finding a positron signal
in the beam at distance $L$ is \cite{yuval}
\beq \label{simsec}
P_{e\mu}^N(x)=\sin^2\theta_{pd} + \sin2\theta_{md} \,
\sin2\theta_{mp} \, \sin^2x.
\eeq
Here $x \equiv {\Delta m^2 L / 4 E}$ and we used $E_1-E_2 \approx
(m_1^2-m_2^2)/2E$, where $E$ is the average energy.  In the limit of
the \SM\ with massive neutrinos ($\theta_{pd}=0$ and
$\theta_{mp}=\theta_{md} \equiv \theta$) eq.~(\ref{simsec}) reduces to
the standard vacuum oscillation probability
\beq
P_{e\mu}(x)=\sin^2x \, \sin^2 2\theta.
\eeq
However, the upper bound (\ref{terange}) implies that $\sin^2 x \le
\O(10^{-4})$ for LSND. Therefore the oscillation part in
(\ref{simsec}) is only a negligible contribution to the required
transition probability (\ref{LSNDdata}) leading to
\beq \label{npimp}
P_{e\mu}^{LSND} = \sin^2\theta_{pd}.
\eeq
We learn that the only significant source for the signal seen at LSND
is a non-vanishing $\theta_{pd} \ne 0$, namely, the produced
(anti)neutrinos are not orthogonal to those that are searched for.  We
note that from experiments we know that neutrino interactions are
dominantly those of the \SM. Therefore, while $\theta_{mp}$ and
$\theta_{md}$ may be large, $\theta_{pd}$ has to be small implying
that the above appearance probability (\ref{npimp}) that arises only
from new interactions must be small.

We first consider new physics effects in purely leptonic interactions. 
(New physics effects in semi-leptonic processes are studied in 
section~\ref{semileptonic}). Such effects are only relevant for the DAR,
where they modify the muon decay. The detection process
is given by the \SM\ and therefore is sensitive only to left-handed
neutrinos.  In that case, the effective interaction for the muon decay 
is given by~\cite{Marshak,PDG}
\beq \label{genlan}
{\cal H^\nu}={4 G_F \over \sqrt{2}} \left[(g^V_{LL})^{\a\b}
(\lbar{e_L} \, \gamma^\mu \, \nu_{\a L})(\lbar{\nu_{\b L}} \, 
\gamma_\mu \, \mu_L) +
(g^S_{RR})^{\a\b} (\lbar{e_R} \, \nu_{\a L})(\lbar{\nu_{\b L}} 
\, \mu_R) \right],
\eeq
where the sum over the weak flavor indices $\a, \b=e,\mu,\tau$ is
implicit. In the \SM\, the only non-vanishing coefficient is
$(g^V_{LL})_{e \mu}=1$ leading to the standard muon decay
\beq \label{SMmuonDecay}
\mu^+ \to e^+ \, \nu_e \, \bar \nu_\mu.
\eeq
We define $G_N^{\nu_\ell}$ to be the effective coupling of the
anomalous muon decays
\beq \label{NPmuonDecay}
\mu^+ \to e^+ \, \bar\nu_e \, \nu_\ell,
\eeq
for $\ell=e,\mu,\tau$ respectively.  In terms of the couplings in
eq.~(\ref{genlan}), $G_N^{\nu_\ell}$ satisfies
\beq
\left|{G_N^{\nu_\ell} \over G_F}\right|^2 = 
\left|(g^V_{LL})_{\ell e}\right|^2 + 
{1 \over 4} \left|(g^S_{RR})_{\ell  e}\right|^2.
\eeq
The three processes in~(\ref{NPmuonDecay}) cannot interfere with each
other because they have different final states. Hence the combined
effective coupling for muon decays that produce $\bar \nu_e$ is
\beq
\left| G_N^\nu \right|^2 = \sum_\ell \left| G_N^{\nu_\ell} \right|^2.
\eeq
In terms of $G_N^\nu$ and for $x \to 0$ the appearance probability
becomes~\cite{yuval}
\beq \label{xzero}
P_{e\mu} = \left|{G_N^\nu \over G_F}\right|^2.
\eeq
From eqs.~(\ref{LSNDdata}) and~(\ref{xzero}) we learn that, in order
to explain the LSND result, the effective \NP\ coupling should satisfy
\beq \label{neededeps}
r \equiv \left|\frac{G_N^\nu}{G_F}\right|^2 =
(3.1^{+1.1}_{-1.0} \pm 0.5) \times 10^{-3}.
\eeq
Thus, at the $90\%\,$C.L. we need
\beq \label{gbound}
r > 1.6 \times 10^{-3}, \qquad 
G_N > 4.0 \times 10^{-2}~G_F.
\eeq
In the next section we study the experimental bounds on $r$.


\section{Experimental Constraints}
\label{expbounds}

The anomalous muon decay (\ref{NPmuonDecay}) is tightly connected to
other lepton flavor violating processes. The \SM\ neutrinos form
$SU(2)_L$ doublets together with the charged left-handed leptons.  As
we will show in Section~\ref{modelind} any theory which gives rise to
the four-Fermi operators that induce the anomalous muon decay
(\ref{genlan}) also necessarily produces the $SU(2)_L$ related
operators of the form
\beq \label{genlal} 
{\cal H^\ell}={4 G_V^\ell \over \sqrt{2}} 
(\lbar{e_L} \, \gamma^\mu \, \ell_L)(\lbar{e_L} \, \gamma_\mu \, \mu_L)
 + {8 G_S^\ell \over \sqrt{2}} 
(\lbar{e_R} \, \ell_L) (\lbar{e_L} \, \mu_R). 
\eeq 
Here $G_V^\ell$ ($G_S^\ell$) are the effective \NP\ vector (scalar)
four-Fermi couplings.  Furthermore we define the combined coupling
\beq
|G_N^\ell|^2 \equiv |G_V^\ell|^2 + |G_S^\ell|^2,
\eeq
for $\ell=e,\mu,\tau$, respectively.  In general, there might be other
interaction terms where all the charged fermions are right-handed.
Clearly, such interactions do not relate to those involving
neutrinos. We therefore ignore such terms and assume that there is no
fine-tuned cancellation between these terms and those we are
considering.

The operators in (\ref{genlal}) mediate \lfvg\ processes involving
only charged leptons.  As we shall see the effective couplings
$G_N^{\nu_\ell}$ and $G_N^\ell$ are always correlated. There is no
experimental evidence for any non-vanishing $G_N^\ell$, so the upper
bounds on $G_N^\ell$ can be used to derive constraints on
$G_N^{\nu_\ell}$. Specifically, the most stringent upper bounds on
$G_N^\ell$ come from $\mue$ (for $\ell=e$), muonium-antimuonium
conversion (for $\ell=\mu$) and $\taumue$ (for $\ell=\tau$).  Before
we turn to a discussion of the exact relation between $G_N^{\nu_\ell}$
and $G_N^\ell$ we present the current experimental bounds on these
three \lfvg\ processes and their implication on $G_N^\ell$.

Using the upper bound $\BR(\mue)< 1.0\times 10^{-12}$ together with
$\BR(\mu~\to~e\,\bar\nu_e\,\nu_\mu)=1$ \cite{PDG} we obtain
\beq \label{muebound}
G_N^e \equiv G_{\mue} < 1.0 \times 10^{-6}~G_F.  
\eeq 
The current bound on the muonium-antimuonium conversion effective
interaction is \cite{psi98}
\beq \label{PSIbound}
G_N^\mu \equiv G_{\mmbar} < 3.0 \times 10^{-3}~G_F.
\eeq 
The upper bound $\BR(\taumuec)< 1.5 \times 10^{-6}$ together with
$\BR(\tau^- \to \mu^- \, \bar\nu_\mu \, \nu_\tau)=0.174$~\cite{PDG}
implies that
\beq \label{taumuebound}
G_N^\tau \equiv G_{\taumue} < 2.9 \times 10^{-3}~G_F.  
\eeq
Note that for \NP\ interactions that have a different Dirac structure
than those in the \SM\ there exist additional constraints on
$G_S^\ell$ which come from the bounds on the Michel
parameters~\cite{PDG}.  We find that at $90\%\,$C.L.
\beq 
G_S^\ell < 3.3 \times 10^{-2}~G_F,
\eeq
which is less severe than the bounds in (\ref{muebound}),
(\ref{PSIbound}) and (\ref{taumuebound}).

If $SU(2)_L$ breaking effects are negligible then $G_N^\ell$ equals to
$G_N^{\nu_\ell}$ up to a factor of at most two from a possible
Clebsch-Gordon coefficient.  If we assume moreover that either
$G_N^\mu$ or $G_N^\tau$ are close to their experimental limit (we will
show later that relaxing these assumptions does not modify our
conclusions) then the experimental bounds (\ref{muebound}),
(\ref{PSIbound}) and (\ref{taumuebound}) imply that
\beq \label{expbound}
G_N^\nu < 6.0 \times 10^{-3}~G_F.
\eeq
Comparing with (\ref{gbound}) we find that in the $SU(2)_L$ symmetric
limit new interactions cannot have a significant contribution to the
DAR signal observed at LSND.

We shall now argue that $SU(2)_L$ breaking effects are in general
small and therefore cannot sufficiently weaken the above bound
(\ref{expbound}).  The crucial ingredient we used to establish
(\ref{expbound}) is $SU(2)_L$ invariance, i.e., we assumed that there
is an $SU(2)_L$ rotation which transforms the four fermion operator
that gives rise to \amd\ to the one where the neutrinos are
replaced by their charged lepton partners.  If $SU(2)_L$ is an exact
symmetry, then the coefficient of both operators coincide (up to a
Clebsch-Gordon factor).  While this relation is exact only when
$SU(2)_L$ is unbroken, from electroweak precision data it follows that
the breaking is small.  As we will discuss in much detail, in the
underlying theory the two related operators are induced by the
exchange of heavy particles, that are members of one $SU(2)_L$
multiplet. If the intermediate particle is a singlet, or if the two
processes are mediated by the same member of the multiplet, then
$G_N^{\nu_\ell}=C_{CG} G_N^\ell$ ($C_{CG}$ is the Clebsch-Gordon
factor).  If not, then the equality is violated and the ratio of
couplings is given by
\beq \label{Mratio}
{G_N^\nu \over G_N^\ell} = C_{CG} {M_1^2 \over M_2^2},
\eeq
where $M_1$ and $M_2$ are the masses of the particles belonging to the
$SU(2)_L$ multiplet that mediate the processes described by $G_N^\ell$
and $G_N^\nu$, respectively.  Then, if $M_1 \ne M_2$ this multiplet
will contribute to the $\rho$ parameter. Thus, we can use the bound on
$\rho-1$ from the electroweak precision measurements, to determine the
maximal ratio in eq.~(\ref{Mratio}).

The contribution to the $\rho$ parameter depends on the Lorentz and
$SU(2)_L$ representation of the multiplet. In general, higher
dimensional representations contribute more. Therefore, it is
sufficient to examine the case of a scalar $SU(2)_L$ doublet, where
the maximal mass splitting can occur.  From the recent data one finds
that at $90\%\,$C.L.~\cite{Pierce}
\beq \label{LEPbound}
\Delta M^2 \equiv |M_1^2-M_2^2| < (77\,{\rm GeV})^2.
\eeq
The mass of the lightest component of any (non-singlet) multiplet is
known to be more than $m_Z/2$ from the measurement of the $Z$
width. Therefore, the largest possible effect arises for $M_2 =
45\,$GeV. Then, the upper bound in~(\ref{LEPbound}) implies that $M_1
< 90\,$GeV and we conclude that
\beq \label{splitbound}
{G_N^\nu \over G_N^\ell} = {M_1^2 \over M_2^2} < 4.0\,.
\eeq
We remark that this is a very conservative estimate for the maximal
value of $M_1^2 / M_2^2$, which could probably be improved by more
rigorous arguments.  Still, it is sufficient to show that the
relaxation of the bound~(\ref{expbound}) due to $SU(2)_L$ breaking
effects could be {\it at most} a factor of four leading to
\beq \label{expboundimprove}
G_N^\nu < 2.4 \times 10^{-2}~G_F, \qquad r < 5.8 \times 10^{-4}.
\eeq
Thus, comparing with (\ref{gbound}) which requires $r > 1.6 \times
10^{-3}$ we learn that the anomalous muon decays \amd\ cannot
significantly contribute to the LSND DAR result even for maximal
$SU(2)_L$ breaking.


\section{Specific models}
\label{specmodel}

In this section we study the mechanism by which heavy intermediate
particles can induce the new four-Fermi interactions. To be specific,
we shall first introduce the general idea within two well-known
extensions of the \SM\ and postpone a model-independent discussion
until the next section.

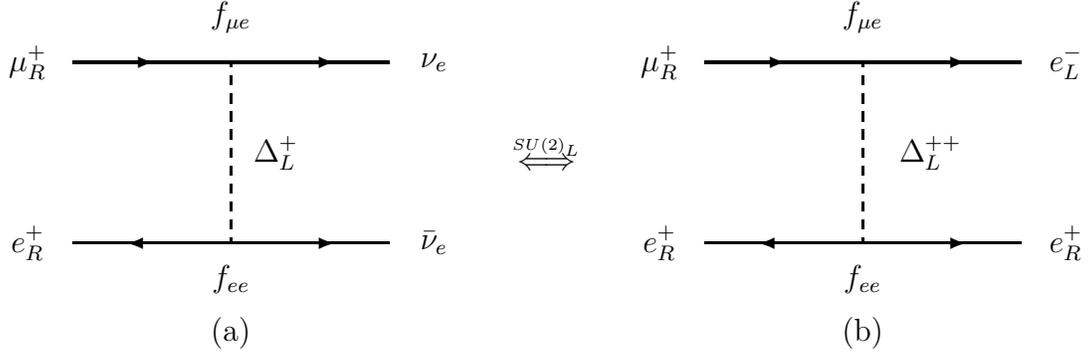
\begin{figure}[htb]
\setlength{\unitlength}{\piclength}
\begin{center}
\begin{picture}(\picwidth,\picheightLRSM)(0,-1)
\thicklines
\multiput(2,2)(14,0){2}
{
 \put(0,0){\line(1, 0){7}} 
 \put(1.2,0){\vector(-1,0){0}} 
 \put(5.8,0){\vector(1,0){0}} 
 \put(3.5,0){\dashbox{0.2}(0,4)}
 \put(0,4){\line(1, 0){7}} 
 \put(1.8,4){\vector(1,0){0}} 
 \put(5.8,4){\vector(1,0){0}} 
}

\put(10,2){\makebox(0,0){$\bar \nu_e$}}
\put(1,6){\makebox(0,0){$\mu^+_R$}}
\put(1,2){\makebox(0,0){$\e^+_R$}}
\put(10,6){\makebox(0,0){$\nu_e$}}
\put(6.5,4){\makebox(0,0){$\Delta_L^+$}}
\put(5.5,7){\makebox(0,0){$f_{\mu e}$}}
\put(5.5,1.2){\makebox(0,0){$f_{e e}$}}
\put(5.5,0){\makebox(0,0){(a)}}

\put(12.5,4){\makebox(0,0){$\SUL$}}

\put(15,2){\makebox(0,0){$\e^+_R$}}
\put(15,6){\makebox(0,0){$\mu^+_R$}}
\put(24,2){\makebox(0,0){$\e^+_R$}}
\put(24,6){\makebox(0,0){$\e^-_L$}}
\put(21,4){\makebox(0,0){$\Delta_L^{++}$}}
\put(19.5,7){\makebox(0,0){$f_{\mu e}$}}
\put(19.5,1.2){\makebox(0,0){$f_{e e}$}}
\put(19.5,0){\makebox(0,0){(b)}}

\end{picture}
\caption{Diagrams for (a) $\mu^+ \to e^+ \, \bar\nu_e \, \nu_e$ 
         and (b) $\mu \to 3e$ for LRSMs \label{nueLR} }
\end{center}
\end{figure}

\begin{figure}[htb]
\setlength{\unitlength}{\piclength}
\begin{center}
\begin{picture}(\picwidth,\picheightLRSM)(0,-1)
\thicklines
\multiput(2,2)(14,0){2}
{
 \put(0,0){\line(1, 0){7}} 
 \put(5.8,0){\vector(1,0){0}} 
 \put(3.5,0){\dashbox{0.2}(0,4)}
 \put(0,4){\line(1, 0){7}} 
 \put(1.8,4){\vector(1,0){0}} 
 \put(5.8,4){\vector(1,0){0}} 
}

\put(3.2,2){\vector(-1,0){0}} 
\put(1,2){\makebox(0,0){$\e^+_R$}}
\put(1,6){\makebox(0,0){$\mu^+_R$}}
\put(10,2){\makebox(0,0){$\bar \nu_e$}}
\put(10,6){\makebox(0,0){$\nu_\mu$}}
\put(6.5,4){\makebox(0,0){$\Delta_L^+$}}
\put(5.5,7){\makebox(0,0){$f_{\mu \mu}$}}
\put(5.5,1.2){\makebox(0,0){$f_{e e}$}}
\put(5.5,0){\makebox(0,0){(a)}}

\put(12.5,4){\makebox(0,0){$\SUL$}}

\put(17.8,2){\vector(1,0){0}} 
\put(15,2){\makebox(0,0){$\e^-_L$}}
\put(15,6){\makebox(0,0){$\mu^+_R$}}
\put(24,2){\makebox(0,0){$\e^+_R$}}
\put(24,6){\makebox(0,0){$\mu^-_L$}}
\put(21,4){\makebox(0,0){$\Delta_L^{++}$}}
\put(19.5,7){\makebox(0,0){$f_{\mu \mu}$}}
\put(19.5,1.2){\makebox(0,0){$f_{e e}$}}
\put(19.5,0){\makebox(0,0){(b)}}

\end{picture}
\caption{Diagrams for (a) $\mu^+ \to \e^+ \, \bar\nu_e \, \nu_\mu$ 
         and (b) $\mu^+ \e^- \to \mu^- \e^+$ for LRSMs \label{numuLR} }
\end{center}
\end{figure}
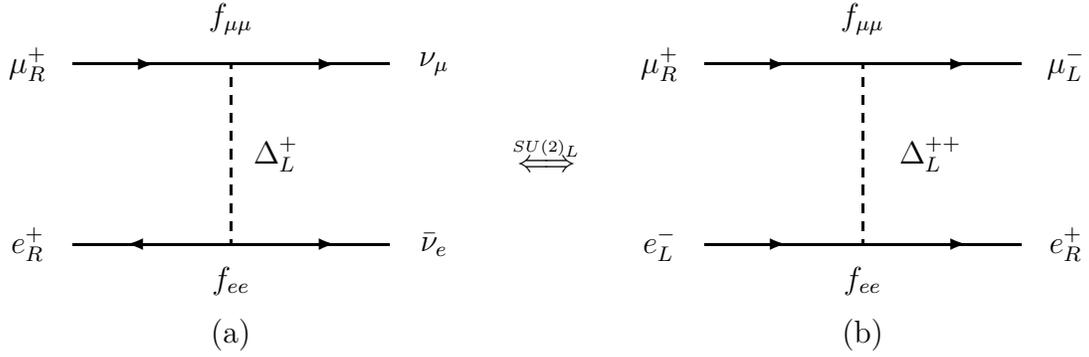

First, we consider the minimal left-right symmetric model 
(LRSM)~\cite{LRS}. The relevant ingredient for our discussion is the
existence of a Higgs triplet, $\Delta_L$, with the following \lfvg\
couplings~\cite{GGMKO}
\beq
{\cal H}_\Delta = i \sum_{\a, \b = e, \mu, \tau} f_{\a \b} 
\left(L_{\a L}^T  \, C \btau_2 \, \Delta_L \, L_{\b L} \right) + 
\rm h.c.~,
\label{tripletLR}
\eeq
where $L_\alpha$ denotes lepton doublet, $C$ is the charge conjugation
matrix and
\beq
\bDelta_L = \pmatrix{\Delta_L^+ / \sqrt{2} & \Delta_L^{++} \cr
                     \Delta_L^0            & -\Delta_L^+ / \sqrt{2}}.
\eeq
$\Delta_L^+$ exchange leads to the effective four fermion interaction
in eq.~(\ref{genlan}) with \cite{Moh,Coarasa,Herczeg}
\beq \label{GnuLR}
|G_N^{\nu_\mu}|  = {|f_{ee} f_{\mu \mu}^*| \over 
2\sqrt{2} M^2(\Delta_L^+)}, \qquad
|G_N^{\nu_\ell}| = {|f_{ee} f_{\mu \ell}^*| \over 
4\sqrt{2} M^2(\Delta_L^+)},
\eeq
where in this case $\ell=e,\tau$ only.  On the other hand, $\Delta_L^{++}$
exchange leads to the related interaction involving four charged
fermions, that we discussed in Section~\ref{expbounds}. The effective
couplings are~\cite{Moh}
\beq \label{GellLR}
|G_N^{\ell}| = {|f_{ee} f_{\mu \ell}^*| \over 
4\sqrt{2} M^2(\Delta_L^{++})}.
\eeq
The diagrams that induce $G_N^{\nu_\ell}$ and $G_N^\ell$ are shown in
Figs.~\ref{nueLR}--\ref{nutauLR}, for $\ell=e, \mu, \tau$,
respectively.  Provided that the mixing of $\Delta_L$ with other Higgs
fields can be neglected, the triplet masses are related via~\cite{Moh}
$M^2(\Delta_L^{++}) + M^2(\Delta_L^{0}) = 2 M^2(\Delta_L^{+})$
implying that $M^2(\Delta_L^{++})/M^2(\Delta_L^{+}) \le 2$.  Then,
using eqs.~(\ref{GnuLR}) and~(\ref{GellLR}) and the bounds
from~(\ref{muebound}), (\ref{PSIbound}) and (\ref{taumuebound}) we
obtain
\beq \label{rdLRS}
r_{\rm LRS} < 1.4 \times 10^{-4}.
\eeq
Thus, even for maximal mass-splitting, within LRSMs \amd\ does not
affect any of the existing terrestrial experiments and, in particular,
cannot explain the LSND DAR-result.

The second example is a supersymmetric extension of the \SM\ without
{$R$-parity} \cite{Howie}.  The trilinear $L_\imath L_\jmath
E_\kappa^c$ couplings between the leptonic chiral superfields $L$ and
$E$ allow \lfvg\ interactions which are mediated by sleptons.  The
relevant couplings are given by
\beq
{\cal H}_{\not R_p} = {\lambda_{\imath \jmath \kappa} \over 2}
\left[\tilde\nu_L^\imath \lbar{\ell_R^\kappa} \ell_L^\jmath 
+\tilde\ell_L^\jmath \lbar{\ell_R^\kappa} \nu_L^\imath
+\tilde\ell_R^{\kappa *} \lbar{\nu_L^\imath}^c \ell_L^\jmath 
-(\imath \to \jmath )\right]+{\rm h.c.} \, ,
\label{lepSS}
\eeq
where $\tilde\nu_L^\imath$ and $\tilde\ell_L^\imath$ denote,
respectively, the sneutrino and the (left-handed) slepton field of
generation $\imath$, and the charge-conjugate fields are defined by
$\nu_R^{\imath c}=C\bar\nu_L^{\imath T}$. In this model \amd\ proceeds
via $\tilde\ell'_L$ exchange~\cite{HaMa}, where $\ell'=\tau~(\mu)$ for
$\ell=\e, \mu~(\tau)$, with effective coupling
\beq 
|G_N^{\nu_\ell}| =
{|\lambda_{1 \ell' 2}\lambda_{\ell \ell' 1}^*|\over 4\sqrt{2} 
M^2(\tilde\ell'_L)}.  
\eeq
On the other hand, the charged lepton processes are mediated by
$\tilde\nu_\tau$ ($\tilde\nu_\mu$) for $\ell=\e, \mu~(\tau)$ with
\beq
|G_N^\ell|=
{|\lambda_{1 \ell' 2}\lambda_{\ell \ell' 1}^*|\over 4\sqrt{2} 
M^2(\tilde\nu_{\ell'})}.
\eeq
The different diagrams that induce $G_N^{\nu_\ell}$ and $G_N^\ell$ are
shown in Figs.~\ref{nueSS}--\ref{nutauSS}, for $\ell=e, \mu, \tau$,
respectively.  We find that $G_N^{\nu_\ell} / G_N^\ell =
M^2(\tilde\ell_L') / M^2(\tilde\nu_{\ell'})$.  The sleptons masses are
related by~\cite{Howie} $M^2(\tilde\ell_L') - M^2(\tilde\nu_{\ell'}) =
m^2_{\ell'} - m_Z^2(1-\sin^2\theta_W) \cos 2 \beta$. Since $\cos
2\beta<0$, in general $M^2(\tilde\ell_L') > M^2(\tilde\nu_{\ell'})$.
Therefore, within SUSY without $R$-parity a possible mass splitting
would only strengthen the $SU(2)_L$ symmetric bound given by
\beq \label{rdRp}
r_{\rm \not R_p} \le 9 \times 10^{-6}.
\eeq
Obviously this is much too small to affect any of the existing
terrestrial experiments and below the range (\ref{neededeps}) needed
to explain the LSND result. As we shall see in the next section the
two explicit examples we presented here in fact exhaust all the
possible purely leptonic couplings induced by intermediate scalar
particles.


\section{Model independent analysis}
\label{modelind}

We have seen in the previous section within two explicit models the
tight relation between the operators that induce \amd\ and those where
the neutrinos are replaced by their charged lepton partners.  In this
section we show in a {\it model independent} way that in our case it
is impossible to evade the close relation between these operators.
 
The exchange of a (heavy) boson between two fermion bilinears induces
a four-fermion operator whose effective coupling at low energies
depends only on the boson mass and the elementary (trilinear)
couplings.  To obtain the most general set of such operators for \NP\
in the leptonic sector only, let us write all trilinear couplings
involving at least one doublet $L$ (which contains the required
neutrino) and at most one singlet $E$ (which contains a right-handed
charged lepton) and the respective antiparticles~\cite{CuDa}.  A
priori there are only four such bilinears to which the intermediate
particle can couple. They are tabulated in Tab.~1 together with their
$SU(2)_L$ representations and the possible values for the charge $Q$
and the hypercharge $Y$ (without loss of generality we suppress the
complex conjugated bilinears which have opposite $Y, Q$).

\begin{center} 
Tab.~1: Lepton-Lepton Bilinears \\
\vspace{0.5cm}
\begin{tabular}{| c | c | c | c | c |} 
\hline  
~Bilinear~ & ~Coupling~ & ~$SU(2)_L$~ & $Q$             & $Y$      \\
\hline \hline  
$L L$      & ~scalar~   & 1, 3        & ~0, $-1$, $-2$~ & $-1$     \\
\hline 
$\bar E L$ & ~scalar~   & 2           & 1, 0            & ~$1/2$~  \\
\hline
$\bar L L$ & ~vector~   & 1, 3        & 1, 0, $-1$      &   0      \\
\hline
$E L$      & ~vector~   & 2           & $-1$, $-2$      & ~$-3/2$~ \\
\hline
\end{tabular}
\end{center} 
\label{LLbilin}

\spa 
Due to the conservation of $Y$ one can only construct operators that
result from the coupling between any of these bilinears and its
complex conjugate. So there are only six possibilities, which we will
discuss one by one: An intermediate scalar singlet cannot contribute
to \amd.  The reason is that the final state of this muon decay has to
contain an $e^+$ and a $\bar \nu_e$. Since the $L L$ bilinear has to
form an $SU(2)_L$ singlet, it has to be antisymmetric in flavor space.
This implies that one cannot produce an $e^+$ and a $\bar \nu_e$
simultaneously by exchanging a charged scalar singlet. (Note that
e.g. $\mu^+ \to \e^+ \, \bar \nu_\tau \, \nu_e$ could be mediated by a
scalar singlet, but that the effective operator responsible for this
process cannot be related by an $SU(2)_L$ rotation to the one where
the neutrinos are replaced by their charged lepton partners.)  The two
remaining possibilities that involve intermediate scalar particles are
those that appeared within the two specific models that we discussed
in the previous section, i.e., the triplet $\bDelta_L$ in LRSMs and
the doublet $L^T=(\tilde \nu_{\ell'}, \tilde \ell')$ in SUSY without
$R$-parity.  We only used model-specific ingredients to explain why in
these models the mass splitting is always smaller than the maximally
allowed. Still, it can be easily checked that even for maximal
splitting the possible effect is too small.  We therefore conclude
that scalar particles in general cannot mediate \amd\ at a rate
required to explain the DAR-result of LSND.

The remaining entries in the Tab.~1 require an intermediate spin-1
boson with vector couplings.  An $SU(2)_L$ singlet couples to $(\bar L
L')_s=\lbar{\nu_\ell} \gamma_\mu \nu_{\ell'} + \lbar{\ell_L}
\gamma_\mu \ell_L'$. This implies that the couplings for any
interaction mediated by a singlet remain exactly the same when
exchanging the two neutrinos by their charged $SU(2)_L$ partners.
Therefore we can directly apply the bound (\ref{expbound}) that we
derived in Section~\ref{expbounds}.  The other option is to have an
$SU(2)_L$ triplet $W'^\mu$ that couples to $\bar L \gamma_\mu \btau L$
just like the \SM\ vector-boson $W$.  If we allow for flavor
off-diagonal couplings $g_{\alpha \beta}$ the exchange of the charged
components (${W'}^\pm$) induces $(\lbar{\nu_e} \gamma_\mu \mu_L) \,
(\lbar{e_L} \gamma^\mu \nu_{\ell})$ and the exchange of the neutral
component ($Z'$) gives rise to $(\lbar{e_L} \gamma_\mu \mu_L) \,
(\lbar{e_L} \gamma^\mu \ell_L)$. For both operators the effective
couplings are proportional to $g_{e\mu} \, g_{e \ell}^*$ and differ
only by the mass splitting between the ${W'}^\pm$ and $Z'$. Thus we
find again that the muon decay \amd\ is tightly related to the charged
lepton decays or muonium-antimuonium conversion.  Finally the $E L$
bilinear requires a spin-1 vector doublet with $Y=3/2$. In this case a
rotation between $\ell$ and $\nu_\ell$ goes along with the exchange of
the two components of this vector doublet, so the $SU(2)_L$ symmetry
guarantees that \amd\ and the respective charged lepton processes have
the same couplings up to the mass-splitting between the two members of
the vector doublet which is small.

We thus conclude that any purely leptonic process, that conserves
total lepton number, cannot contribute significantly to the LSND
DAR-signal.


\section{Semi-leptonic interactions}
\label{semileptonic}

So far we have restricted our analysis to the case of having \NP\ only
in \amd\ showing that its rate cannot be sufficient to provide $\bar
\nu_e$'s at a rate seen at LSND. While this is a reasonable assumption
for LRSMs, where the intermediate particles that induce the new
interactions only couple to leptons, in general also new semi-leptonic
interactions can play a role.

In fact, for LSND the production reaction for the DIF ($\pi^+ \to
\mu^+ \, \nu$) and the detection reaction of both the DIF ($\nu \, n
\to p \, e^-$) and the DAR ($\bar \nu \, p \to n \, e^+$) are
semi-leptonic.  All the semi-leptonic four-Fermi operators of
relevance to LSND involve a $u$ and a $d$-quark, a charged lepton and
only {\it one} neutrino. While the involved quarks necessarily belong
to the first generation and the charged leptons must be either the
muon or the electron, a priori all the three neutrino flavors could be
involved in the \NP\ contribution to the semi-leptonic reactions.

The four-Fermi operators that are relevant for the detection reactions
are of the form $(\nu_\ell e^+ d \bar u)$ and for the DIF production
the relevant operator is $(\nu_\ell \mu^+ d \bar u)$.  Applying
similar arguments as in our discussion of the purely leptonic new
interactions, one can use the $SU(2)_L$ symmetry to relate these
operators to the ones where the neutrino is replaced by its charged
lepton partner, namely
\beqs 
(\nu_\ell e^+   d \bar u) &\SUL& (\ell^- e^+   q \bar q) 
\label{slff1} \\
(\nu_\ell \mu^+ d \bar u) &\SUL& (\ell^- \mu^+ q \bar q)
\label{slff2},
\eeqs
where $q=u, d$.

Let us ignore for the moment $SU(2)_L$ breaking effects and the Dirac
structure that we suppressed in~(\ref{slff1}) and~(\ref{slff2}).  (It
is more complicated than for the purely leptonic case and we will
discuss how these operators arise and relate at the elementary level
later on.) Then the upper bounds on processes which would be induced
by the operators that contain only charged particles can be used to
put stringent constraints on the semi-leptonic reactions relevant to
LSND.
 
For $\ell=\mu$ in~(\ref{slff1}) and $\ell=e$ in~(\ref{slff2}) the
strongest constraint comes from the bounds on muon conversion on
nuclei~\cite{PDG}
\beq
{\sigma(\mu^- {\rm Ti} \to e^- {\rm Ti}) \over
\sigma({\rm all}~\mu^- {\rm Ti}~{\rm capture})} < 4.3 \times 10^{-12}.
\eeq
For the effective coupling of the ($\bar\mu_L \gamma_\mu e^+_L 
q_L \gamma^\mu \bar q_L$) operator this implies
\beq \label{gslbound}
G_N(\mu e q q) < 2.1 \times 10^{-6}~G_F,
\eeq
which is four orders of magnitudes smaller than the coupling
in~(\ref{gbound}) needed to produce a signal for LSND.  We note that
the above bound (\ref{gslbound}) could be somewhat relaxed due to
differences in the matrix elements~\cite{Davidson}, different Dirac
structure and $SU(2)_L$ breaking effects which we ignored.  Still,
assuming that there are no fine-tuned cancellations, it is safe to
conclude that the coupling of semi-leptonic reactions which violate
only the $L_e$ and $L_\mu$ lepton family numbers are much too small to
be relevant for LSND.

If the $\nu_\tau$ is involved then all the three lepton family numbers
are violated and new interactions are required for both the production
and detection processes. In this case the relevant experimental bounds
are~\cite{PDG}
\beq
\BR(\tau \to e \pi^0) < 3.7 \times 10^{-6}, \qquad
\BR(\tau \to \mu \pi^0) < 4.0 \times 10^{-6}. \qquad
\eeq
Normalizing these branching ratios to $\BR(\tau^- \to \pi^- \nu_\tau)
= 0.11$ and using isospin symmetry we find that the coupling of the
operator $(\bar\tau_L \gamma_\mu \ell^+_L q_L \gamma^\mu \bar q_L)$,
satisfies the constraint
\beq
G_N(\tau \ell q q) <  8.5 \times 10^{-3}~G_F,
\eeq
for $\ell = e, \mu$.  For the DIF, $SU(2)_L$ relates this coupling to
those describing the production~($\ell=\mu$) and the
detection~($\ell=e$) process. For the DAR, the production must be
\amdtau. Using the agreement between the tau lifetime and its purely
leptonic decay width, one can conclude that BR(\amdtau) $< 5 \times
10^{-3}$. Therefore, complications that arise from isospin breaking
effects, possible different Dirac structure and $SU(2)_L$ breaking 
effects can be safely ignored also in this case.  
We conclude that the constraints arising from muon conversion on nuclei and 
$\tau \to \ell \pi^0$ exclude new semi-leptonic interactions from
significantly affecting either of the two LSND results.

We turn now to a model independent analysis of the possible couplings
and their relations for the semi-leptonic channels using similar
arguments as in Section~\ref{modelind}. Our goal is to show explicitly
that it is impossible to evade the tight relation between operators
related by $SU(2)_L$ rotations, which is crucial for the arguments
presented above to be valid in general.  Consider first the bilinears
that consist of {\it two} quarks. In order to couple to the leptonic
bilinears of Tab.~1 they must be $SU(3)_C$ singlets.  Hence they
contain one quark $Q, D$ or $U$ and one anti-quark, where $Q$ is the
doublet and $D$ and $U$ are $SU(2)_L$ singlets. They are summarized in
Tab. 2.

\begin{center} 
Tab.~2: Quark-Quark Bilinears \\
\vspace{0.5cm}
\begin{tabular}{| c | c | c | c | c |} 
\hline  
~Bilinear~ & ~Coupling~ & ~$SU(2)_L$~ & $Q$              &  $Y$   \\
\hline \hline  
$\bar U Q$ & ~scalar~   & 2           & 0, $-1$          & ~$-1/2$~ \\
\hline
$\bar D Q$ & ~scalar~   & 2           & 1, 0             &  1/2   \\
\hline
$\bar Q Q$ & ~vector~   & 3, 1        & 1, 0, $-1$       &   0    \\
\hline
$\bar U D$ & ~vector~   & 1           & $-1$             &  $-1$    \\
\hline
$\bar U U$ & ~vector~   & 1           & 0                &   0    \\
\hline
$\bar D D$ & ~vector~   & 1           & 0                &   0    \\
\hline
\end{tabular} 
\end{center} 
\label{QQbilin}
\spa 

Due to the conservation of $Y$ it follows that from the
singlet-singlet bilinears (the last three entries in Tab.~2) only 
$\bar U U$ and $\bar D D$ couple to the vector singlet of $\bar
L L$ of Tab.~1. However the resulting four fermion operators do not
contribute to the semi-leptonic reactions of interest since they
cannot change the charge of the involved leptons and quarks.

The $\bar U Q$ and $\bar D Q$ bilinears couple via a scalar $SU(2)_L$
doublet (with $Y=\pm 1/2$) to $\bar E L$ (or its complex
conjugate). Let us use here the familiar notation from SUSY without
$R$-parity for the couplings and the scalar particles (none of our
arguments requires supersymmetry and therefore the underlying theory
providing the new couplings is arbitrary).  The coupling
$\lambda'_{\imath \jmath \kappa} L_\imath Q_\jmath
D_\kappa^c$ between the chiral superfields $L$, $Q$ and
$D$ induces exactly the required coupling between the quark bilinear
and the scalar doublet
\beq
\lambda'_{\imath \jmath \kappa} 
\left[ \tilde\nu_L^\imath \lbar{d_R^\kappa} d_L^\jmath -
       \tilde\ell_L^\imath \lbar{d_R^\kappa} u_L^\jmath  
\right] + {\rm h.c.} \, .
\label{lqd}
\eeq
The coupling of the scalar doublet to the lepton bilinear proceeds via
the appropriate term in (\ref{lepSS}). Obviously the presence of the
charged scalar doublet member $\tilde\ell_L^\imath$, which mediates
the semi-leptonic processes relevant to LSND, generically requires the
presence of its neutral doublet partner $\tilde\nu_L^\imath$. Then the
effective couplings for the operator $(\lbar{u_L} \, d_R) \,
(\lbar{\ell'_R} \, \nu_\ell)$ ($\ell \ne \ell'$) coincides with the
effective couplings for the operator $(\lbar{d_L} \, d_R) \,
(\lbar{\ell'_R} \, \ell_L)$ up to the mass splitting between
$\tilde\ell_L^\imath$ and $\tilde\nu_L^\imath$. However, as we have
shown in the beginning of this section this operator is severely
constrained and thus cannot significantly contribute to semi-leptonic
processes that change lepton flavor.  So we conclude that an
intermediate scalar doublet cannot contribute significantly to the
LSND results.

The remaining candidate for a coupling to a leptonic bilinear is the
$\bar Q Q$.  In this case the intermediate particle must be either a
spin-1 triplet or singlet with $Y=0$. Since the singlet is neutral it
cannot mediate the charged-current semi-leptonic processes that we are
interested in.  For the triplet we require a vector boson $W'$ that
has both flavor diagonal couplings to quarks and flavor off-diagonal
couplings $g_{\ell \ell'}$ to leptons (like the one we evoked for the
self-coupling of the $\bar L L$ bilinear).  Then the exchange of the
charged components (${W'}^\pm$) induces $(\lbar{u_L} \gamma_\mu d_L)
\, (\lbar{\ell_L} \gamma^\mu \nu_{\ell'})$ and the exchange of the
neutral component ($Z'$) gives rise to $(\lbar{q_L} \gamma_\mu q_L) \,
(\lbar{\ell_L} \gamma^\mu \ell'_L)$. For both operators the effective
couplings are proportional to $g_{\ell \ell'}$ and differ only by the
mass splitting between the ${W'}^\pm$ and $Z'$.  Thus the argument
using related processes containing only charged fermions that we
presented in the beginning works equally well for an intermediate
spin-1 boson.

Having exhausted the quark-quark bilinears we now turn to the
possibility of having bilinears containing both a lepton and a quark
that couple to leptoquarks \cite{Davidson}. At least one
bilinear must contain the doublet $L$ (since we require a neutrino)
and any of the quark fields $Q, D$ and $U$ leading to the following
combinations:

\begin{center} 
Tab.~3a: Quark-Lepton $L$ Bilinears \\
\vspace{0.5cm}
\begin{tabular}{| c | c | c | c | c |} 
\hline  
~Bilinear~ & ~Coupling~ & ~$SU(2)_L$~ & $Q$                   &   $Y$  \\
\hline \hline  
$Q L$      & ~scalar~   & 1, 3        & ~2/3, $-1/3$, $-4/3$~ & ~$-1/3$~ \\
\hline 
$\bar D L$ & ~scalar~   & 2           & 1/3, $-2/3$           &  $-1/6$  \\
\hline
$\bar U L$ & ~scalar~   & 2           & $-2/3$, $-5/3$        &  $-7/6$  \\
\hline
$\bar Q L$ & ~vector~   & 1, 3        & 1/3, $-2/3$, $-5/3$   &  $-2/3$  \\
\hline 
$D L$      & ~vector~   & 2           & $-1/3$, $-4/3$        &  $-5/6$  \\
\hline
$U L$      & ~vector~   & 2           & 2/3, $-1/3$           &   1/6  \\
\hline
\end{tabular} 
\end{center} 
\label{bilinears}

\spa
\noindent
Let us first consider those four-Fermi operators that are built only
from the bilinears of Tab.~3a.  The first three bilinears in Tab.~3a
require scalar couplings, while the other three can only couple to a
spin-1 particle.  The conservation of angular momentum forbids that
bilinears that have a different type of couplings couple to each
other.  Then using the conservation of hypercharge one can see that
the allowed four-fermion operators only arise from bilinears that
couple to themselves.  It follows that operators from bilinears
with a singlet quark always contain either the $D$ or the $U$
singlet quarks, but not both, and are therefore of no relevance
to the semi-leptonic reactions that could explain LSND.

The $Q L$ bilinear could couple either to a scalar singlet or triplet
of $SU(2)_L$. The singlet coupling involves the term $u_L \, \ell_L -
d_L \nu_\ell$ implying that a vertex where the scalar singlet couples
to the neutrino and the $d$-quark has the same coupling strength as to
the charged lepton partner and the $u$-quark. Thus the operator
$(\lbar{u_L} \, \lbar{\ell_L}) \, (d_L \, \nu_{\ell'})$ ($\ell \ne
\ell'$) has the same coupling as $(\lbar{u_L} \, \lbar{\ell_L}) \,
(u_L \, \ell'_L)$ and we can again apply our argument using the bounds
on $\mu$~Ti~$\to e$~Ti and $\tau \to \ell \pi^0$.  Similarly, when the
$Q L$ bilinear forms an $SU(2)_L$ triplet it is the $Q=1/3$ component
that is relevant which couples to $u_L \, \ell_L + d_L \nu_\ell$.
Again, the same arguments as for the singlet case apply, and the
related charged lepton processes put severe bounds on this case as
well.

Thus the only remaining candidate is the $\bar Q L$ bilinear.  The
intermediate leptoquark $X$ must be a spin-1 boson.  It could be a
$SU(2)_L$ singlet with $Q=Y=2/3$ that induces the operator
$(\lbar{\nu_\ell} \gamma_\mu u_L + \lbar{\ell_L} \gamma_\mu d_L) \,
(\lbar{u_L} \gamma^\mu \nu_{\ell'} + \lbar{d_L} \gamma^\mu \ell'_L)$,
which obviously gives the same couplings for the $SU(2)_L$ related
processes that we study.  An intermediate spin-1 triplet $X^\mu$ with
$Y=2/3$ couples to $\bar Q \gamma_\mu \btau L$.  The relevant
coupling for our discussion is induced by the $Q=2/3$ component of
$X_\mu$ which couples via $\tau_3$ to the fermions
$\sum_{\ell=e, \mu, \tau} h_\ell X_\mu^{(2/3)} 
\left[\lbar{u_L} \gamma^\mu \nu_\ell - \lbar{d_L} \gamma^\mu \ell_L \right]$. 
Hence -- no surprise -- the operator $(\lbar{u_L} \gamma_\mu
\nu_\ell) \, (d_L \gamma^\mu \lbar{\ell'_L})$ has the same effective
coupling as $(\lbar{d_L} \gamma_\mu \ell_L) \, (d_L \gamma^\mu
\lbar{\ell'_L})$ and the discussed $SU(2)_L$ symmetry also works for this
case.

Finally, we have to consider the case when the bilinears from Tab.~3a
do not couple to themselves but to those containing a lepton singlet
$E$ and a quark field. The possible bilinears are given in Tab.~3b.

\begin{center} 
Tab.~3b: Quark-Lepton $E$ Bilinears \\
\vspace{0.5cm}
\begin{tabular}{| c | c | c | c | c |} 
\hline  
~Bilinear~ & ~Coupling~ & ~$SU(2)_L$~ & $Q$              &   $Y$  \\
\hline \hline  
$\bar Q E$ & ~scalar~   & 2           & $-2/3$, $-5/3$   &  $-7/6$  \\
\hline 
$D E$      & ~scalar~   & 1           & $-4/3$           &  $-4/3$  \\
\hline
$U E$      & ~scalar~   & 1           & $-1/3$           &  $-1/3$  \\
\hline
$Q E$      & ~vector~   & 2           & ~$-1/3$, $-4/3$~ & ~$-5/6$~ \\
\hline 
$\bar D E$ & ~vector~   & 1           & $-2/3$           &  $-2/3$  \\
\hline
$\bar U E$ & ~vector~   & 1           & $-5/3$           &  $-5/3$  \\
\hline
\end{tabular} 
\end{center} 
\label{bilinQE}

\spa
Comparing the entries for the bilinears in Tab.~3a and
Tab.~3b we find that there are four possibilities:
$\bar Q E$ and $U \bar L$ via a scalar doublet, 
$U E$ and $\bar Q \bar L$ via a scalar singlet, 
$Q E$ and $\bar D \bar L$ via a vector doublet and
$\bar D E$ and $Q \bar L$ via a vector singlet. 
Repeating similar arguments as presented before one can show 
that in all of these cases also the corresponding charge lepton
operators are induced.

We thus conclude that lepton number conserving semi-leptonic processes
cannot contribute significantly to any of the LSND signals.


\section{Implications and Conclusions}
\label{conclusion}

Extensions of the \SM\ in general do not conserve individual lepton
numbers and therefore provide an alternative mechanism for neutrino
flavor conversion that may show up in neutrino oscillation
experiments.  We have argued that the experimental constraints on such
\lfvg\ interaction do not allow such an interpretation for any of the
LSND results. Our argument relies on the bounds from $SU(2)_L$ related
reactions containing the charged partner of the relevant neutrino. We
have shown explicitly the relations between the effective coupling of
the two types of reactions within LRSMs and SUSY without $R$-parity as
examples for \NP\ that affects the anomalous muon decay.  Moreover, we
have demonstrated in a model-independent way that the ratio of these
couplings is always of order one and that the deviation from unity is
only due to a generically small mass splitting between the bosonic
members of an $SU(2)_L$ multiplet and some Clebsch-Gordan
coefficients.

It is still interesting to ask whether \lfv\ could influence other
neutrino experiments and whether their explanation in terms of
neutrino oscillation might be modified or even spoiled by the \NP.
Recall that both the solar and atmospheric neutrino experiments detect
quite a large deviation from the predicted neutrino-flux by a factor
$\sim 1/2$ with experimental uncertainties of about 10\%. In general
the effects of \NP\ on the production or detection process are much
smaller and hence cannot influence those experiments drastically via
those processes.  However, if the MSW effect is the correct solution
to the \SN\ problem, then \NP\ may influence the resonant
conversion \cite{nMSW} if reactions of the type
\beq \label{NPsun}
\nu_e f \to \nu_\ell f, 
\eeq
where $f=e, u, d$ and $\ell=\mu$ or $\tau$, are present.  Note that
while the process (\ref{NPsun}) and the flavor violating semi-leptonic
reactions that we discussed always violate the individual lepton
numbers $L_e$ and $L_\ell$ by {\it one} unit, this is only true for
the anomalous muon decay $\mu^+ \to \e^+ \, \bar \nu_e \, \nu_e$. The
two other decays producing $\nu_\mu$ or $\nu_\tau$ in the final state
violate $L_e$ by {\it two} units. Hence the \NP\ processes that are
potentially relevant for short baseline experiments and the MSW
mechanism are not necessarily related.  But generically all types of
reactions could be present. While saturating the current bounds on the
effective couplings for reactions involving a $\nu_\tau$ is not
sufficient to produce a significant effect for LSND (since they are
suppressed both for the production {\it and} the detection), this is
not true for \SN s. A detailed analysis \cite{Bergmann} shows that in
this case the region in the parameter space that corresponds to the
small mixing angle solution is somewhat shifted. The shift is
basically in the value of the mixing angle, while the required
mass-squared difference is almost unaltered by the presence of the
\NP. We note that also the effect of \lfvg\ interactions on the
resonant neutrino conversion in supernovae has been
studied~\cite{Mansour,BergmannKagan} with the result that here one can
have drastic changes to the neutrino survival probability for a large
region of parameter space.

We conclude that the presence of \lfvg\ interactions cannot solve the
problem of explaining the three observed $\Delta m^2$ scales with
three neutrino generations. We did not study the possibility that 
lepton number violation processes may be relevant. For example,
the decay \amdlv\ may explain the DAR LSND result \cite{Herczeg}. 

\acknowledgements
We thank Haim Goldberg, Yossi Nir, Damien Pierce, Tom Rizzo,
Tom Weiler and Jim Wells for helpful discussions.
Y.G. is supported by the U.S. Department of Energy under contract
DE-AC03-76SF00515.




\newpage

\hspace{\picdist}


\hspace{\picdist}
\begin{figure}[htb]
\setlength{\unitlength}{\piclength}
\begin{center}
\begin{picture}(\picwidth,\picheightLRSM)(0,-1)
\thicklines
\multiput(2,2)(14,0){2}
{
 \put(0,0){\line(1, 0){7}} 
 \put(1.2,0){\vector(-1,0){0}} 
 \put(3.5,0){\dashbox{0.2}(0,4)}
 \put(0,4){\line(1, 0){7}} 
 \put(5.8,0){\vector(1,0){0}} 
}

\put(7.8,6){\vector(1,0){0}} 
\put(3.8,6){\vector(1,0){0}} 
\put(1,2){\makebox(0,0){$\e^+_R$}}
\put(1,6){\makebox(0,0){$\mu^+_R$}}
\put(10,2){\makebox(0,0){$\bar \nu_e$}}
\put(10,6){\makebox(0,0){$\nu_\tau$}}
\put(6.5,4){\makebox(0,0){$\Delta_L^+$}}
\put(5.5,7){\makebox(0,0){$f_{\mu \tau}$}}
\put(5.5,1.2){\makebox(0,0){$f_{e e}$}}
\put(5.5,0){\makebox(0,0){(a)}}

\put(12.5,4){\makebox(0,0){$\SUL$}}

\put(21.2,6){\vector(-1,0){0}} 
\put(17.2,6){\vector(-1,0){0}} 
\put(15,2){\makebox(0,0){$\e^+_R$}}
\put(15,6){\makebox(0,0){$\mu^-_L$}}
\put(24,2){\makebox(0,0){$\e^+_R$}}
\put(24,6){\makebox(0,0){$\tau^+_R$}}
\put(21,4){\makebox(0,0){$\Delta_L^{++}$}}
\put(19.5,7){\makebox(0,0){$f_{\mu \tau}$}}
\put(19.5,1.2){\makebox(0,0){$f_{e e}$}}
\put(19.5,0){\makebox(0,0){(b)}}

\end{picture}
\caption{Diagrams for (a) $\mu^+ \to \e^+ \, \bar\nu_e \, \nu_\tau$ 
         and (b) $\taumuec$ for LRSMs \label{nutauLR} }
\end{center}
\end{figure}


\newpage
\begin{figure}[htb]
\setlength{\unitlength}{\piclength}
\begin{center}
\begin{picture}(\picwidth,\picheightSUSY)(0,0)
\thicklines
\multiput(2,2)(14,0){2}
{
 \put(0,0){\line(1, 1){2}} \put(0.8,0.8){\vector(-1,-1){0}} 
 \put(0,4){\line(1,-1){2}} \put(1.2,2.8){\vector(1,-1){0}} 
 \put(2,2){\dashbox{0.2}(3,0)}
 \put(5,2){\line(1, 1){2}} \put(6.2,3.2){\vector(1, 1){0}} 
 \put(5,2){\line(1,-1){2}} \put(6.2,0.8){\vector(1,-1){0}} 
}

\put(1,1){\makebox(0,0){$\bar \nu_e$}}
\put(1,7){\makebox(0,0){$\mu^+_L$}}
\put(10,1){\makebox(0,0){$\nu_e$}}
\put(10,7){\makebox(0,0){$\e^+_L$}}
\put(2,4){\makebox(0,0){$\lambda_{132}$}}
\put(9,4){\makebox(0,0){$\lambda_{131}$}}
\put(5.5,5){\makebox(0,0){$\tilde \tau_L$}}
\put(5.5,1){\makebox(0,0){(a)}}

\put(12.5,4){\makebox(0,0){$\SUL$}}

\put(15,1){\makebox(0,0){$\e^+_R$}}
\put(15,7){\makebox(0,0){$\mu^+_L$}}
\put(24,1){\makebox(0,0){$\e^-_L$}}
\put(24,7){\makebox(0,0){$\e^+_L$}}
\put(16,4){\makebox(0,0){$\lambda_{132}$}}
\put(23,4){\makebox(0,0){$\lambda_{131}$}}
\put(19.5,5){\makebox(0,0){$\tilde \nu_\tau$}}
\put(19.5,1){\makebox(0,0){(b)}}

\end{picture}
\caption{Diagrams for (a) $\mu^+ \to e^+ \, \bar\nu_e \, \nu_e$ 
         and (b) $\mu \to 3e$ for SUSY without $R_p$ \label{nueSS} }
\end{center}
\end{figure}
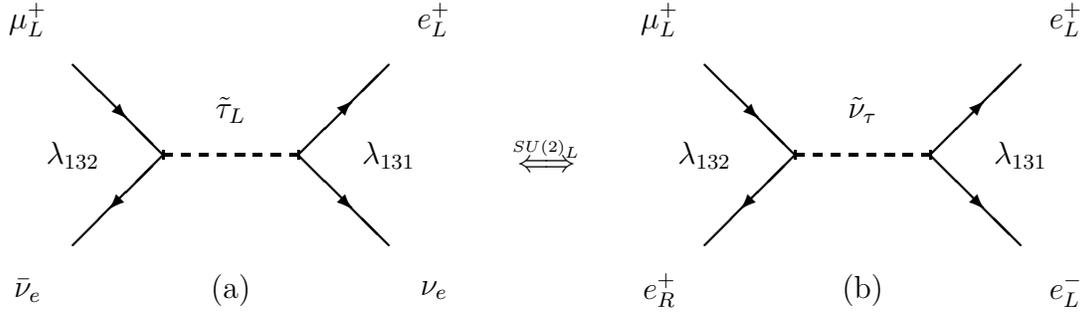


\hspace{\picdist}
\begin{figure}[htb]
\setlength{\unitlength}{\piclength}
\begin{center}
\begin{picture}(\picwidth,\picheightSUSY)(0,0)
\thicklines
\multiput(2,2)(14,0){2}
{
 \put(0,4){\line(1,-1){2}} \put(1.2,2.8){\vector(1,-1){0}} 
 \put(2,2){\dashbox{0.2}(3,0)}
 \put(5,2){\line(1, 1){2}} \put(6.2,3.2){\vector(1, 1){0}} 
 \put(5,2){\line(1,-1){2}} \put(6.2,0.8){\vector(1,-1){0}} 
}

\put(2,2){\line(1, 1){2}} \put(2.8,2.8){\vector(-1,-1){0}} 
\put(1,1){\makebox(0,0){$\bar \nu_e$}}
\put(1,7){\makebox(0,0){$\mu^+_L$}}
\put(10,1){\makebox(0,0){$\nu_\mu$}}
\put(10,7){\makebox(0,0){$\e^+_L$}}
\put(2,4){\makebox(0,0){$\lambda_{132}$}}
\put(9,4){\makebox(0,0){$\lambda_{231}$}}
\put(5.5,5){\makebox(0,0){$\tilde \tau_L$}}
\put(5.5,1){\makebox(0,0){(a)}}

\put(12.5,4){\makebox(0,0){$\SUL$}}

\put(16,2){\line(1, 1){2}} \put(17.2,3.2){\vector(1,1){0}} 
\put(15,1){\makebox(0,0){$\e^-_L$}}
\put(15,7){\makebox(0,0){$\mu^+_L$}}
\put(24,1){\makebox(0,0){$\mu^-_L$}}
\put(24,7){\makebox(0,0){$\e^+_L$}}
\put(16,4){\makebox(0,0){$\lambda_{132}$}}
\put(23,4){\makebox(0,0){$\lambda_{231}$}}
\put(19.5,5){\makebox(0,0){$\tilde \nu_\tau$}}
\put(19.5,1){\makebox(0,0){(b)}}

\end{picture}
\caption{Diagrams for (a) $\mu^+ \to \e^+ \, \bar\nu_e \, \nu_\mu$ 
         and (b) $\mu^+ \e^- \to \mu^- \e^+$ for SUSY 
         without $R_p$ \label{numuSS} }
\end{center}
\end{figure}
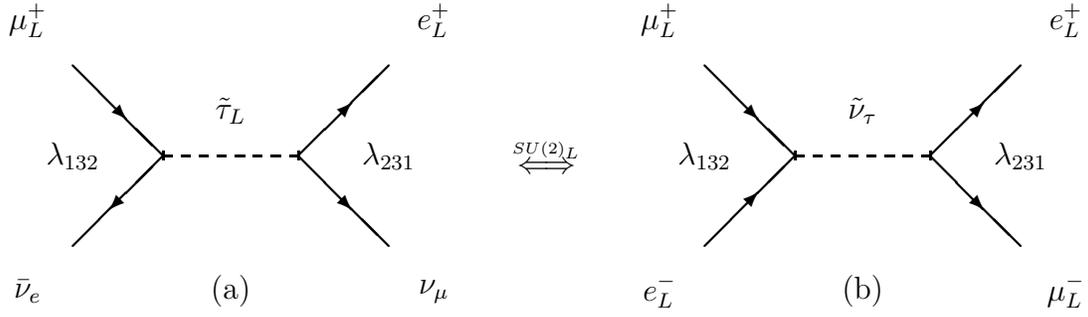


\hspace{\picdist}
\begin{figure}[htb]
\setlength{\unitlength}{\piclength}
\begin{center}
\begin{picture}(\picwidth,\picheightSUSY)(0,0)
\thicklines
\multiput(2,2)(14,0){2}
{
 \put(0,0){\line(1, 1){2}} \put(0.8,0.8){\vector(-1,-1){0}} 
 \put(2,2){\dashbox{0.2}(3,0)}
 \put(5,2){\line(1, 1){2}} \put(6.2,3.2){\vector(1, 1){0}} 
}

\put(2,6){\line(1,-1){2}} \put(3.2,4.8){\vector(1,-1){0}} 
\put(7,4){\line(1,-1){2}} \put(8.2,2.8){\vector(1,-1){0}} 
\put(1,1){\makebox(0,0){$\bar \nu_e$}}
\put(1,7){\makebox(0,0){$\mu^+_L$}}
\put(10,1){\makebox(0,0){$\nu_\tau$}}
\put(10,7){\makebox(0,0){$\e^+_L$}}
\put(2,4){\makebox(0,0){$\lambda_{122}$}}
\put(9,4){\makebox(0,0){$\lambda_{321}$}}
\put(5.5,5){\makebox(0,0){$\tilde \mu_L$}}
\put(5.5,1){\makebox(0,0){(a)}}

\put(12.5,4){\makebox(0,0){$\SUL$}}

\put(16,6){\line(1,-1){2}} \put(16.8,5.2){\vector(-1,1){0}} 
\put(21,4){\line(1,-1){2}} \put(21.8,3.2){\vector(-1,1){0}} 
\put(15,1){\makebox(0,0){$\e^+_R$}}
\put(15,7){\makebox(0,0){$\mu^-_R$}}
\put(24,1){\makebox(0,0){$\tau^+_R$}}
\put(24,7){\makebox(0,0){$\e^+_L$}}
\put(16,4){\makebox(0,0){$\lambda_{122}$}}
\put(23,4){\makebox(0,0){$\lambda_{321}$}}
\put(19.5,5){\makebox(0,0){$\tilde \nu_\mu$}}
\put(19.5,1){\makebox(0,0){(b)}}

\end{picture}
\caption{Diagrams for (a) $\mu^+ \to \e^+ \, \bar\nu_e \, \nu_\tau$ 
         and (b) $\taumuec$ for SUSY 
         without $R_p$ \label{nutauSS} }
\end{center}
\end{figure}


\end{document}